# Evaluation of research activities of universities of Ukraine and Belarus: a set of bibliometric indicators and its implementation

*Vladimir Lazarev, Serhii Nazarovets and Alexey Skalaban*

Monitoring bibliometric indicators of University rankings is considered as a subject of a University library activity.

In order to fulfill comparative assessment of research activities of the universities of Ukraine and Belarus the authors introduced a set of bibliometric indicators. A comparative assessment of the research activities of corresponding universities was fulfilled; the data on the leading universities are presented. The sensitivity of the one of the indicators to rapid changes of the research activity of universities and the fact that the other one is normalized across the fields of science condition advantage of the proposed set over the one that was used in practice of the corresponding national rankings.

## 1. Bibliometric Indicators of University Rankings

The use of bibliometric data for developing rankings of activities of not only "purely" academic organizations but also of the universities and other institutions of higher education is reviewed in a special section of the Scientometrics Manual (Akoev et al., 2014). As it is noted therein, "in international University rankings such as Academic Ranking of World Universities <…>, <…>THE WUR and QS World University Ranking, the indicators related to publication activity contain from 20 to 60 % of the final measured score that proves the importance of scientific publications for the evaluation of the University — both its educational and academic, as well as international component" (Akoev et al., 2014). Moreover, the Leiden Ranking is being constituted of the indicators that are exclusively various data on the citedness of publications created at a University and various bibliometric data on such publications themselves. It is also stated that a whole range of possible bibliometric indicators, "starting with mere number of publications and finishing with a number of normalized indicators of citedness" are used "for the correct account of the impacts of specific universities" of natural-scientific and technical profile (Akoev et al., 2014). It is but natural because "there are appealing issues in bibliometrics, among them <…> the multicriterion evaluation of actors, especially universities" (Zitt, 2005).

## 2. Bibliometric Monitoring of University Research Activity as an Objective of a University Library

Monitoring bibliometric indicators of University rankings and even inventing such indicators might be recognized as an objective of a University library because very similar indicators are used in more traditional professional operations of libraries of this kind. Such a viewpoint seems to be even more justified if there is not any special unit for assessment of University research activity. Therefore, the practice of bibliometric assessment of University research activity by University libraries is quite typical for a number of the Eastern European countries.

In the opinion of the authors of the present paper (who all were affiliated with University libraries themselves when the paper was in preparation) such an activity is fairly inherent for University libraries and other academic libraries not only because of their familiarity with bibliometrics. Since a library is a significant structural division of a University, it must play an increasingly important role in the activities of a modern University by performing some functions that were previously uncharacteristic for libraries at all. We believe that only if University libraries *variously* facilitate universities in addressing the main challenges that universities face, they really demonstrate their relevance in a modern society. One of such challenges is improving a University's position in the world rankings, and the Scientific Library of the Belarusian National Technical University has a certain experience in facilitating such an improving (Skalaban, 2013), while the Academic Library of the National University of "Kyiv-Mohyla Academy" does its best to stimulate scientific publication activity of the University staff (Nazarovets, 2016). So, as for creating or amending bibliometric sets of indicators that reflect effectiveness of research activity of a University, this seems to be an even more obvious objective for a competent University or academic librarian. Also, our University libraries – both Ukrainian and Belarusian gained an experience to fulfill bibliometric studies of the efficiency of research activities of our universities (Borisova, 2016; Skalaban, Yurik & Lazarev, 2017).

## 3. Bibliometric Rankings of Universities of Ukraine and Belarus

Bearing the above-stated in mind, one might agree that it is quite natural to compare the scientific performance of universities of the countries of the former Soviet Union by bibliometric indicators. An example of such practice is the Ukrainian "Ranking of Universities according to the Scopus Indicators" (see http://osvita.ua/vnz/rating/51053/ ) Another example is the "Ranking of Educational Institutions of the Republic of Belarus and of Scientific-and-Research Institutions of the Educational Institutions by H-Index, SCOPUS Database" that was being prepared in 2012- 2016 at the Central Scientific Library of the National Academy of Sciences of Belarus (CSL) and was updated on the library website (http://csl.bas-net.by/Web/Pages/Periodicals/pdf/scopus-vuz.pdf). Creating their own bibliometric rankings was caused both by the interest to bibliometrics as an evaluation tool and by insufficient presence of Belarusian and Ukrainian universities in the world most popular rankings. Thus, e.g., in Sept 2016 only two Belarusian universities, viz. Belarusian State University and Belarusian National Technical University were presented in the QS Ranking (see http://www.bsu.by/main.aspx?guid=146761). A representatives of a lot of countries, indeed, might feel that their universities are insufficiently presented in the main world ranking systems because "although most systems claim to produce rankings of *world* universities, the analysis of *geographical coverage* reveals substantial differences between the systems as regards the distribution of covered institutions among geographical regions. It follows that the systems define the 'world' in different manners, and that — compared to the joint distribution of the five systems combined — each system has a proper orientation or bias, namely *U-Multirank* towards Europe, *ARWU* towards North America, *Leiden ranking* towards emerging Asian countries, and *QS* and *THE* towards Anglo-Saxon countries" (Moed, 2017).

As it was stated above, the corresponding Belarusian bibliometric ranking has been designed and was being maintained by staff members of an academic library, viz. of the CSL. In the ranking having been compiled by the CSL since 2012 to 2016 the indicators were: the number of publications of the organization, as reflected in Scopus database; the number of citations recorded in the Scopus database to the publications of the organization; the H-index. The H-index rating was clearly considered as the main one (that was reflected not only in the very title of the ranking, but also in the paper devoted to the ranking (Berezkina, Sikorskaya & Khrenova, 2013). However, by its very nature, the H-index "cannot diminish over time, <...> and a scientist might have many years to stay retired and not to write scientific works, while his H-index would not be less than it was at the height of his career" (Akoev et al., 2014). Similarly, a University might occupy a high place in a ranking due to its past scientific advances. "Therefore, in order to obtain a more meaningful measure one should use a publication window as in case with any bibliometric magnitude <...>. For example, all the articles published <...> over a five-year period may be considered, and citations obtained by these articles may be taken into account" (Akoev et al., 2014). The problems of efficacy of University research that is "driven by assessment and performance targets" (as a consequence of the general problems of "top-down planning and reduced local autonomy for departments") that universities faced in recent decades require rapid assessments of the current state of research activities, but not the cumulative assessment of all achievements that ever occurred (Martin, 2016).

Another restriction of the H-index is the absence of normalization at the disciplinary field level. As it is stated in Scientometrics Manual, "comparison of the absolute values of the index among scientists working in different fields of science is impossible as it is not a field normalized indicator" (Akoev et al., 2014). As Ton van Raan stated, "because the H-index does not take into account the often large differences in citation density between, and even within, fields of science, this indicator is in many situations not appropriate for the assessment of research performance" (Raan, 2013). Therefore it is not by all means reasonable to apply the H-index to researches being fulfilled in various fields and, since that, – at various institutions. But a user of the Ukrainian and Belarusian Rankings would unconditionally compare, say, a food University with a medical University regardless the difference in publication and citation practice in the corresponding disciplinary fields.

Moskaleva (2013) states that bibliometric indicators "applicability depends on the size of the compared samples. If we compare bibliometric indicators of the two organizations working in the same field about the same time period and also comparable in accordance with the number of scientists working at them, then any of these indicators can show the superiority of one of the organizations or their equality. However, if one of the organizations exists for 20 years and the other – for 5, or if they carry out research in different scientific fields, or differ in the number of scientists, none of the indicators directly may not be used, the normalization of differences both in the science fields and in the number of authors <...> is required" (Moskaleva, 2013). After all, in order to make decisions in an organization management one commonly uses fresh data for equal periods of an organization activity, and the very concept of "efficiency" involves consideration of costs, including the salaries of the staff that are obviously different as the staffs of different organization are different in quantity. It is therefore considered that "the size of the organization almost everywhere is taken into account by normalization of differences among the number of faculty staff or academic staff" (Akoev et al., 2014). Thus, bibliometric evaluation of the scientific performance of the organization should be normalized across the fields of science, to relate to the recent period of time and to be normalized at the number of staff level. It is absolutely obvious, and we pay so much attention to these aspects only due to the fact that the above-mentioned conditions were not met in designing the Ukrainian and Belarusian Rankings that both consist of the same indicators: the number of publications of the organization, as reflected in Scopus database; the number of references recorded in the Scopus database to publications of the organization; the H-index. The rankings compilers used the latter as the indicator in accordance to the descending magnitude of which the universities are placed in a ranking list.

## 4. Searched and Found Indicators to Be Used for Bibliometric University Rankings of Ukraine and Belarus Instead of the Discarded Ones

So, which bibliometric indicators should be chosen for the assessment of research efficiency of universities (or any research organizations) as the appropriate ones? "World practice is to use typically two indicators for evaluation a scientist, viz. the total number of citations to his publications and the average number of citations to his publication", – writes I. V. Marshakova-Shaikevich (2013).  As for her own practice, I. V. Marshakova-Shaikevich reports the results of "the research activity of universities of Russia in 2006-2010" on the basis of a number of indicators, three out of them being considered as the most important, viz. the total number of their publications, 2006-2010, as reflected by the InCites™ in the Web of Science™ database; the total number of citations registered in the Web of Science™ to the publications of 2006-2010 and the average number of citations to a document (out of sample of publications of 2006-2010) according to the Web of Science™. These three indicators, in our opinion, should be considered mandatory for the evaluation of efficiency of research activity of an organization because the total number of citations to the publications created at an organization indicates the documented total use of the documentary flow, created at an organization over a period of time, and, indirectly, indicates the value of the cited documentary flow (as *value* is a property of an object that is being cognized through the satisfaction of the desires of human beings that is conditional, in general, on the use of an object); the average number of citations to a publication indicates the use and value of an average publication from the documentary flow and the number of publications themselves indicates *ipso facto* the productivity of the researchers of the institution (Lazarev, 2017). It should be reminded that, if the correlation of the concepts of the value of scientific documents and of scientific performance of an institution, at which they were created, seems to be unqustionable, the relevance of the concepts of productivity of researchers of an organization to efficiency of research activity of the latter is much more contentious. However, when such databases as Web of Science™ and Scopus (practicing very rigid selection of periodicals, articles from which are reflected by them), are used for productivity evaluation, the productivity is considered to be highly selective as relates to articles published in the "highest quality" sources. Thus the productivity data occurred to be selective and just relative; but essentially this is not a disadvantage but rather an advantage, because with this approach, to some extent, the quality of the publications themselves is taken into account: the presence of publication in these databases testified that it has exceeded a certain threshold of the quality of periodicals in which they were published.

As for citations, it is interesting to know both the total number of citations to the publications of an organization and the amount of citations to its average publication. But when evaluating *different* organizations, if it is not possible to carry out data normalization at the differences in the number of their employees, the amount of citations to an *average* publication acquires a key importance as "balancing" the inequality of quantity of received citations that is caused by differences in publication practice determined by a varieties of quantities of contributors working at different organizations. (However, normalization at the differences among the fields of science will not be achieved in this case).

Therefore, out of the three above-stated useful indicators, only the third one occurs to be of *key* significance, viz. the average number of citations to one article, while the first and second ones being rather the "raw material" for its formation. In the paper by Marshakova-Shaikevich (2013) these three indicators were obtained from the Web of Science™; they also can be obtained from the Scopus database.

We believe it is appropriate also to use data on the number of publications of a University authors relating to the 10% most cited ones out of total amount of publications of the same year and of the same research field *as one more indicator of key significance*: we consider them as reflecting the presence of outstandingly excellent researches at a university. In the paper by Bornmann et al. (2015) the presence of the top-cited papers is considered to be a significant separate indicator of "scientific excellence".

The number of publications of a University that belong to the 10% most cited ones, might be obtained by using the SciVal integrated modular platform that analyzes the activities of research organizations based on data from the Scopus. These data are normalized at the level of the fields of science; that is, using these data, along with previous ones, we meet another above-mentioned requirement to a correct bibliometric evaluation of efficiency of research activities of an organization.

For our study the data taken from Scopus (according the state of affairs by September 30, 2016) were used. Taken into account were the indicators of those universities of Belarus and Ukraine that had at least 20 documents included in Scopus during 2011-2015. Thus, we tried to assess the research activities at universities for a specific period, close to the current one, but not their activities since their foundation. The data on the first 10 Belarusian and Ukrainian universities are discussed below.

## 5. Results and Discussion

Table 1 represents data on the "top ten" Ukrainian universities in line with the values of the chosen indicators; they are placed in order of descending values of the "number of publications belonging to the 10% most cited publications of same subjects".

| TABLE 1 Ten Ukrainian universities according to the magnitudes of indicators adopted in the study and calculated with the aid of the Scopus data | | | | |
|---|---|---|---|---|
| University | The number of publications in the 10% most cited publications of same subjects (according to SciVal), value/rank | The average citedness of an article, value/rank | Number of citations, 2011-2015 | Number of publications (articles, reviews), 2011-2015 |
| Taras Shevchenko National University of Kyiv | 297/1 | 2,56/3 | 9003 | 3518 |
| V. N. Karazin Kharkiv National University | 99/2 | 2,29/4 | 3855 | 1685 |
| Ivan Franko National University of Lviv | 85/3 | 2,24/6 | 3040 | 1353 |
| Odessa I.I. Mechnikov National University | 54/4 | 3,18/1 | 1727 | 543 |
| Lviv Polytechnic National University "Lviv Polytechnic" | 44/5 | 1,46/10 | 1109 | 757 |
| Sumy State University | 40/6 | 2,28/5 | 1420 | 622 |
| National Technical University of Ukraine "Igor Sikorsky Kyiv Polytechnic Institute" | 35/7 | 1,65/9 | 1667 | 1011 |
| National Technical University "Kharkiv Polytechnic Institute" | 35/7 | 1,8/8 | 909 | 504 |
| Yuriy Fedkovych Chernivtsi National University | 32/9 | 2,64/2 | 1397 | 530 |
| Tavrida National V.I. Vernadsky University | 19/10 | 1,93/7 | 808 | 419 |

As compared with the data of the "official" Ukrainian Ranking (http://osvita.ua/vnz/rating/51053/) the ranks of the majority of the "top ten" Ukrainian universities remained the same. However, there are some significant differences. Due to the use of the described indicators the Sumy State University and  entered the "top ten" of the universities of Ukraine. This may indicate the intensification of research activities of scientists of these universities in 2011-2015, which was not recorded in the evaluation attempts of the Ukrainian Ranking (http://osvita.ua/vnz/rating/51053/) that were undertaken without regard to the  chronological  framework  and  to  the  presence  of  outstandingly valuable  research results obtained by scientists of these universities.

Despite the small number of publications of scientists of the Odessa I.I. Mechnikov National University, in average, each publication created by its authors was cited three times, and this is the best result among the Ukrainian "top ten" universities. In its turn, a large number of publications of the National Technical University of Ukraine "Igor Sikorsky Kyiv Polytechnic Institute" were cited less often, that can be interpreted as an evidence of need for better representation of the results of researches of the National Technical University scientists. A similar remark seems to be true in respect of the publication activity of scientists of Lviv Polytechnic National University "Lviv Polytechnic".

Table 2 represents corresponding data on the "top ten" Belarusian universities; they are also placed in order of descending values of the "number of publications belonging to the 10% most cited publications of same subjects". (It should be noted that in fact the Table 2 features the 11 universities, as the "top ten universities" determined in accordance with the "number of publications belonging to the 10% most cited publications of same subjects" and in accordance with the "average citation of one article" do not coincide with each other; the variance is one university.)

In general, positions taken by the most of the universities also did not differ much with the ones stated in the "official" Belarusian Ranking (http://csl.bas-net.by/Web/Pages/Periodicals/pdf/scopus-vuz.pdf ) – even taking into account the fact that in the cited Ranking the three-fold reflection of the Belarusian State University took place: as a separate University and as its two affiliated research institutions. However, there are significant differences also in respect of the two universities: the Gomel State Medical University and the Grodno State Medical University have been ranked in our "top ten", but not in the Ranking. Moreover, the Gomel State Medical University, that occupied only the 16[th] place in the rank list developed in accordance with the Hirsch index (the 14[th] one if we consider the three-fold reflection in the ranking of the Belarusian State University), ranked in our list the *first* place by the average citedness per one article of 2011-2015! The magnitudes of the indicators of Table 2 that are attributed to these two universities are indicative of the intensification of research activities of scientists working at them in a recent time period and demonstrate the inadequacy of the H-index to assess the current state of scientific activities of an organization.

| TABLE 2 Eleven Belarusian universities according to the magnitudes of indicators adopted in the study and calculated with the aid of the Scopus data | | | | |
|---|---|---|---|---|
| University | The number of publications of the 10% most cited publications of same subjects (according to SciVal), value/rank | The average citedness of an article, value/rank | Number of citations, 2011-2015 | Number of publications (articles, reviews), 2011-2015 |
| Belarusian State University | 128/1 | 2,42/6 | 3475 | 1435 |
| Belarusian National Technical University | 26/2 | 2,69/3 | 652 | 242 |
| Belarusian State University of Informatics and Radioelectronics | 11/3 | 1,81/9 | 429 | 236 |
| Gomel State Medical University | 7/4 | 4,47/1 | 206 | 46 |
| Grodno State Medical University | 7/4 | 1,91/8 | 213 | 111 |
| Belarusian State Technological University | 7/4 | 1,15 | 204 | 176 |
| Belarusian State Medical University | 6/7 | 2,57/4 | 193 | 75 |
| F. Skorina Gomel State University | 6/7 | 1,70/10 | 318 | 186 |
| Brest State University named after A.S. Pushkin | 4/9 | 3,89/2 | 113 | 29 |
| Sukhoi State Technical University of Gomel | 3/10 | 2,08/7 | 104 | 50 |
| Yanka Kupala State University of Grodno | 1 | 2,45/5 | 228 | 93 |

Let us notice that the Belarusian State University that was the recognized leader according to the Belarusian Ranking (http://csl.bas-net.by/Web/Pages/Periodicals/pdf/scopus-vuz.pdf ) data took only the 6th place according to the magnitude of the average citedness per one article, although the workers of this University published in 2011-2015 the largest amount of articles and reviews (as reflected in the Scopus database). The first and second rank according to the magnitude of the average citedness per one article were respectively received by the Gomel State Medical University and the Brest State University named after A. S. Pushkin, that had published, respectively, 5 and 8 times smaller amount of articles and reviews (as reflected in the Scopus database) than the Belarusian National Technical University that had received the 3rd rank according to the magnitude of the average citedness per one article. These data demonstrate that a large number of publications even in prestigious periodicals do not in the least guarantee a good level of their citedness.

It should be noted that the main obstacle to the carrying out any scientometric analysis of activity of the academic establishments of Belarus and Ukraine is the poor quality of the data presented in the affiliation profiles – both in the databases of the Web of Science™ platform and of the Scopus. For example, the analytical tool SciVal that uses the Scopus data recognizes institutions only of the primary affiliations as fixed in Scopus. If, however, some publications indicated a version of the affiliation title, that differs from its one fixed in the profile, such publications would form a "pseudo-profile" of the Scopus data and, accordingly, such records would not be reflected in the genuine profile of the institution and will not be taken into account when constructing the rankings.

In the prestigious rankings of world universities, such as the Academic Ranking of World Universities, THE WUR, QS World University Rankings, bibliometrics is used together with other indications (survey of experts, the number of teaching staff, level of funding, etc.), presenting a university administrators with enough information on the state of research activities at their institutions along with the other one. The technique that the authors propose in this paper is based solely on the selected bibliometric indicators, which is insufficient for a comprehensive analysis of universities. However, this indicator is believed to meet the requirements of the monitoring the research activities of them.

## 6. Conclusion

Thus, in order to ensure monitoring of the efficiency of research activities of universities of natural-scientific and technical profiles of the Eastern European countries and taking into account their incomplete representation in the leading international rankings we suggested to use of a set of bibliometric indicators, different from that was used in the "Rankings of Educational Institutions of the Republic of Belarus and of Scientific-and-Research Institutions of the Universities by H-Index, SCOPUS Database" (http://csl.bas-net.by/Web/Pages/Periodicals/pdf/scopus-vuz.pdf) and from that also being used in the "Rankings of Universities according to the Scopus indicators"

**Vladimir Lazarev** (*vslazarev@bntu.by*) is the Head of a Unit at the Information Technologies Department of the Scientific Library of the Belarusian National Technical University, Belarus;

**Serhii Nazarovets** (*serhii.nazarovets@gmail.com*) is the Assistant Professor of Kyiv National University of Culture and Arts, Ukraine;

**Alexey Skalaban** (*skalaban@gmail.com*) is an expert of the NEICON (National Electronic Information Consortium), Russia.